\documentclass[conference]{IEEEtran}

\usepackage{amsmath}
\usepackage{graphicx}
\usepackage{cite}
\usepackage{hyperref}

\begin{document}
	
	% 标题和作者信息
	\title{A Demo of Radar Sensing Aided Rotatable Antenna for Wireless Communication System}
	\author{
		\IEEEauthorblockN{Qi Dai\IEEEauthorrefmark{1}, Beixiong Zheng\IEEEauthorrefmark{1}, Qiyao Wang\IEEEauthorrefmark{1}, Xue Xiong\IEEEauthorrefmark{1}, Xiaodan Shao\IEEEauthorrefmark{2}, Lipeng Zhu\IEEEauthorrefmark{3}, Rui Zhang\IEEEauthorrefmark{3}\IEEEauthorrefmark{4}}
		\IEEEauthorblockA{
			\IEEEauthorrefmark{1}School of Microelectronics, South China University of Technology, Guangzhou 511442, China\\
			\IEEEauthorrefmark{2}Department of Electrical and Computer Engineering, University of Waterloo, Waterloo, ON N2L 3G1, Canada\\
			\IEEEauthorrefmark{3}Department of Electrical and Computer Engineering, National University of Singapore, Singapore 117583, Singapore\\
			\IEEEauthorrefmark{4}School of Science and Engineering, The Chinese University of Hong Kong, Shenzhen 518172, China\\
			Email:202321061987@mail.scut.edu.cn; bxzheng@scut.edu.cn;  202420165372@mail.scut.edu.cn; \\ftxuexiong@mail.scut.edu.cn; x6shao@uwaterloo.ca; zhulp@nus.edu.sg; rzhang@cuhk.edu.cn}
	}
	
	% 创建标题
	\maketitle
	
	% 摘要
	\begin{abstract}
	As a new antenna architecture, rotatable antenna (RA) can flexibly adjust each antenna's boresight/orientation to enhance communication performance cost-effectively. In this demonstration, we develop a prototype of radar sensing-aided RA system, where radar sensing module is applied to aid dynamic antenna orientation. The prototype consists of a transmitter (TX) module and a receiver (RX) module, both of which employ universal software radio peripherals (USRPs) for transmitting and receiving signals, respectively. Specifically, the RA-based TX first utilizes a laser radar to detect the RX's location and collects its angle of arrival (AoA) information. Then based on the above information, the antenna servo is controlled by STM32 microcontroller to adjust the deflection angle, which enables the RA to align its boresight direction with the located RX. Experimental results verify the effectiveness of the prototype and indicate that the RA-based system significantly outperforms the conventional fixed-antenna system.
	\end{abstract}
	
	% 关键词
	
	% 介绍部分
	\vspace{-0.15cm}	
	\section{Introduction}
		\vspace{-0.15cm}
	%As wireless communication technologies rapidly advance, the sixth-generation (6G) networks aim to accommodate a vastly larger user group while ensuring superior communication quality, driving a growing demand for more adaptive wireless systems \cite{7397887}. 
	
	Multiple-input-multiple-output (MIMO) is a cornerstone technology that substantially enhances the capacity, reliability, and spectral efficiency of wireless systems. To accommodate diverse emerging applications and infrastructures, the sixth-generation (6G) wireless network is envisioned to deliver superior performance through ultra-high data rate, ultra-low latency, enhanced spectral efficiency, and versatile connectivity. This, in turn, drives the evolution of MIMO technologies into more advanced paradigms, such as massive MIMO and extremely large-scale MIMO\cite{7397887}. However, the aforementioned MIMO technologies are based on fixed-antenna architecture, which not only leads to increased hardware cost and energy consumption, but also fails to adapt to diverse and dynamic communication environments due to the inability to adjust orientation or position of each antenna\cite{lu2014overview}. To overcome this issue, fluid antenna system (FAS) and movable antenna (MA) have been introduced to enhance spatial diversity, beamforming, and multiplexing performance by enabling flexible movement of antennas/arrays \cite{wong2020fluid,zhu2023modeling,zhu2023movableantenna}. To fully exploit the spatial degrees of freedom (DoFs), recent advancements in six-dimensional movable antenna (6DMA) technology allow for flexible adjustment of both three-dimensional (3D) positions and 3D rotations of antennas/arrays, thereby achieving improved communication performance through adaptive allocation of antenna resources \cite{shao20246d,shao20246dma}. While these techniques demonstrate superior performance compared to fixed-antenna, the associated hardware cost and complexity of implementation remain critical factors affecting practical feasibility. As a result, there are still technical challenges to overcome for the practical deployment of FAS/MA/6DMA.
		\begin{figure}[!t]
		\centering
		\includegraphics[width=3.5in]{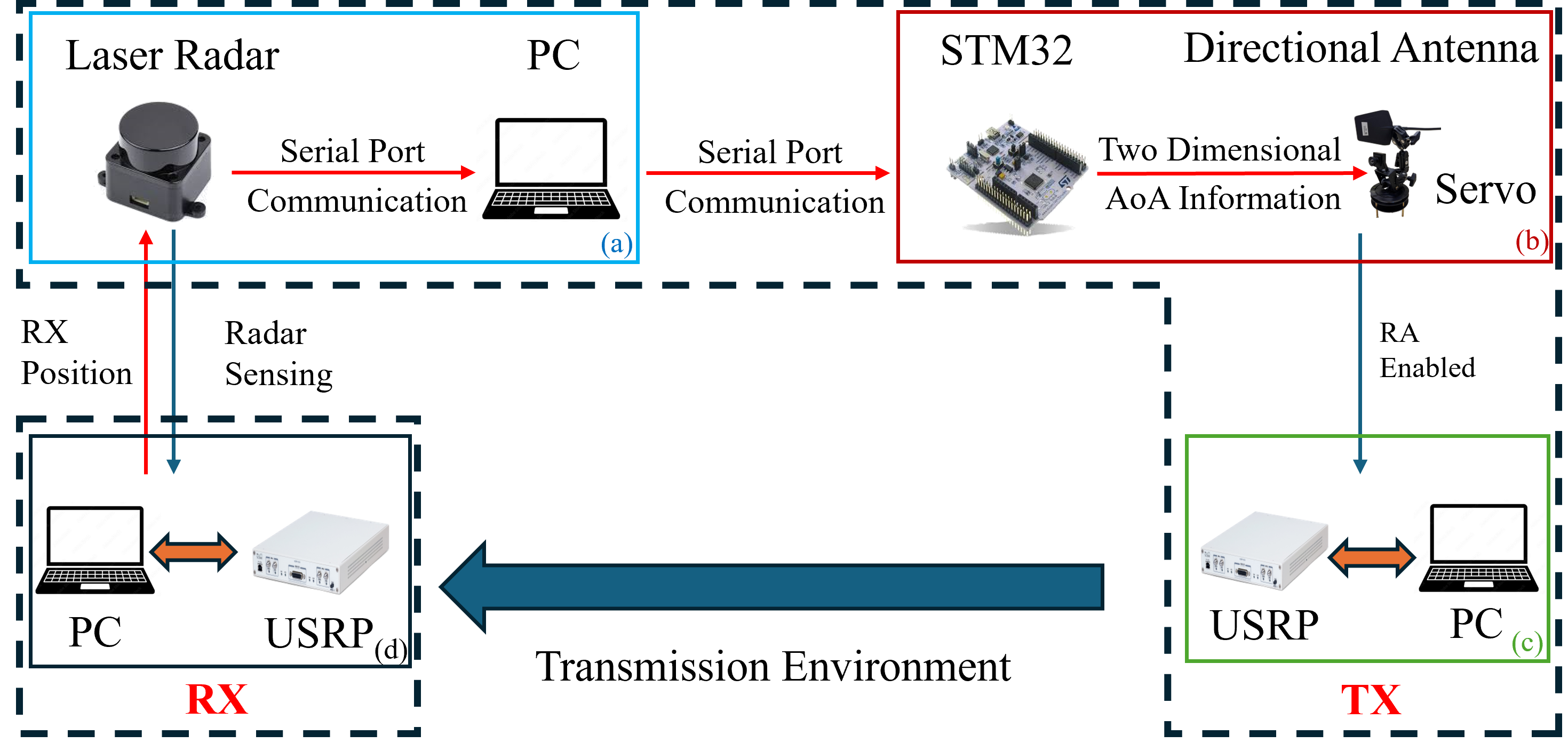} % 替换 "your-image-file" 为图片文件名，例如 "system_overview.png"
		\vspace{-0.5cm}
		\caption{System architecture. (a) Laser radar module. (b) RA module. (c) RA enabled USRP module at TX. (d) USRP module at RX.}
		\label{fig:system}
		\vspace{-0.5cm}
	\end{figure}	
	
		\begin{figure*}[!t] % 使用 figure* 环境
		\vspace{-1.2cm}
		\centering
		\includegraphics[width=5.6in]{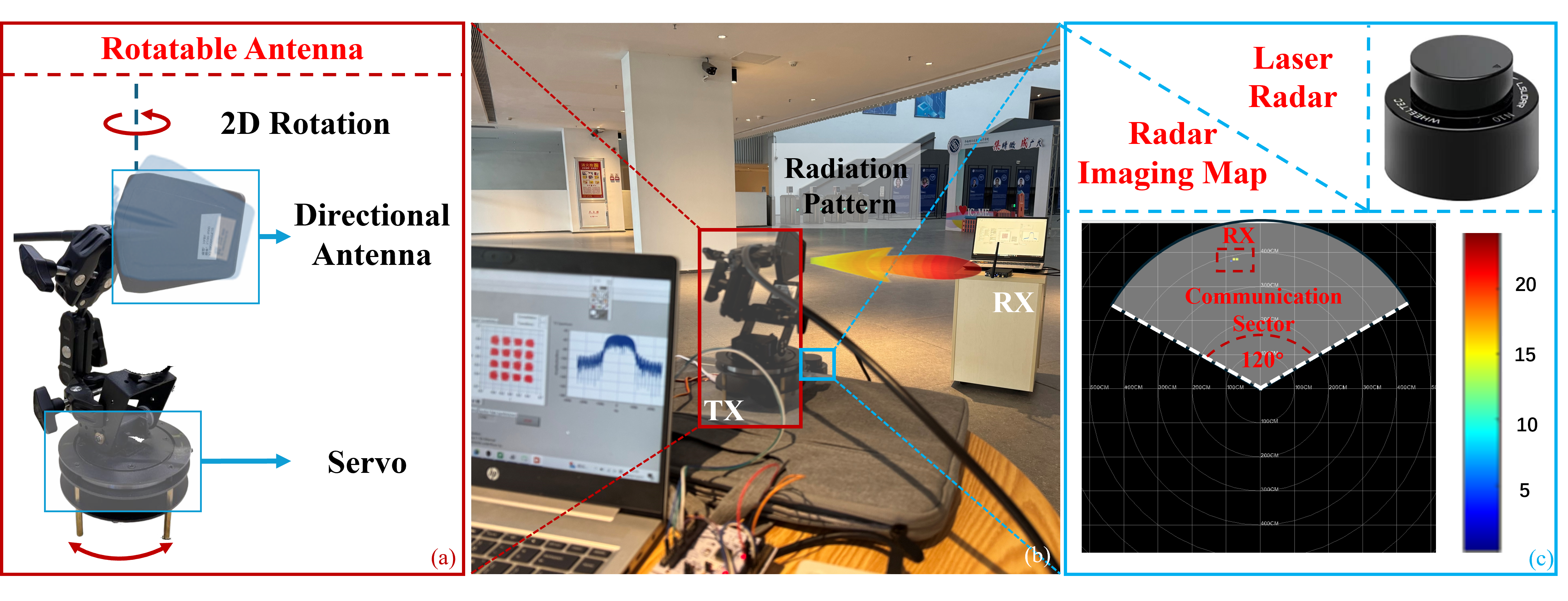} % 图片宽度设置为整栏宽度，或根据需要调整
		\vspace{-0.2cm}
		\caption{(a) Rotatable antenna prototype. (b) Test scenario. (c) Laser radar and radar imaging map.}
		\label{fig:system}
		\vspace{-0.4cm}
	\end{figure*}	
	Motivated by above, rotatable antenna (RA) has emerged as a promising and cost-effective solution for adaptive wireless communication systems\cite{wu2024modelingoptimizationrotatableantenna,zheng2025rotatableantennaenabledwireless}. By enabling dynamic adjustment of individual antenna's deflection angle, RA can flexibly modify its antenna boresight/orientation to attain a desired directional radiation pattern, thereby improving the array gains in specified directions. Such capability allows RA to adapt to various communication environments by exploiting additional spatial DoFs. However, the real-time reconfiguration of antenna boresight/orientation in practical communication systems is a challenging yet interesting problem, which deserves further investigation and validation. Considering the above advantages and design issues of RA, we develop a radar sensing-aided RA prototype to enhance communication coverage and quality. Experimental results are provided to demonstrate the superior communication performance of the RA system to the fixed-antenna system.

			\vspace{-0.20cm}
	\section{System Overview}
		\vspace{-0.14cm}
		
	The architecture of the proposed prototype, as depicted in Fig. 1, consists of two parts: an RA-based transmitter (TX) and an omni-directional dipole antenna-based receiver (RX). Specifically, the TX first utilizes a laser radar to detect the location of the RX. Subsequently, the STM32 microcontroller adjust the servo motor to align the RA's boresight direction with the located RX based on RX's AoA information. The main components of each module are elaborated as follows.

	\subsection{\textit{Transmitter}}

	The TX consists of three primary components: the laser radar module, the RA module, and the universal software radio peripheral (USRP) module.
	
	\hangindent=2.em \hangafter=1 $\bullet$ \textbf{Laser Radar Module} [Fig. 1(a)]: The laser radar module, which consists of a laser radar and a personal computer (PC), is responsible for determining the location of the RX. First, the radar conducts a 360° scan of the surrounding environment or scans within a predefined area, and then determines the location of the RX by leveraging the time-of-flight (TOF) principle. Second, the radar captures and encodes the information in terms of the RX's light intensity, distance, and two-dimensional (2D) AoA into a hexadecimal-formatted data cluster per detection cycle. Finally, these data clusters will be further processed on a PC. In particular, under a radar operating frequency of 10Hz, we calculate the average of AoA information collected over one-second intervals to reduce detection error.
	
	\hangindent=2.0em \hangafter=1 $\bullet$ \textbf{RA Module} [Fig. 1(b)]: The RA module, which compromises a STM32 microcontroller, a servo, and a servo-controlled directional antenna, is responsible for precisely aligning the antenna boresight direction towards the RX according to the radar detection data. Specifically, after receiving the averaged AoA data from laser radar module, the STM32 generates pulse-width modulation (PWM) signals to adjust the servo's angular configuration, ensuring that the directional beam of the antenna is accurately oriented towards the RX. In particular, the module is equipped with a vertically polarized directional antenna that offers a maximum gain of 10 dBi and a beamwidth of 60°. Since the radar module provides only 2D AoA data, adjustment in the zenith angle of the directional antenna is excluded in the current prototype. Additionally, a proportional-integral-derivative (PID) control algorithm is implemented to enhance the alignment accuracy.
	
	\hangindent=2.0em \hangafter=1 $\bullet$ \textbf{USRP Module} [Fig. 1(c)]: The USRP module, which consists of a USRP device and a PC, serves as the transmission endpoint in the communication system. The PC manages the USRP device to conduct key functions such as signal generation, modulation, and parameter configuration. After a specific modulation scheme, such as quadrature amplitude modulation (QAM), is selected, the transmission signal will be encoded and modulated using software tools such as LabVIEW. Then, the USRP transmits the modulated signal to the RX over the wireless channel via RA.

	\subsection{\textit{Receiver}}
	
	As illustrated in Fig. 1(d), the RX is equipped with another USRP, operating within the frequency range from 70 MHz to 6 GHz. Furthermore, the USRP at the RX utilizes an omni-directional dipole antenna and is connected to a PC responsible for data processing and result visualization.

		\vspace{-0.14cm}
    \section{Experimental Results}
	\vspace{-0.10cm}

	In this section, we present the experimental results to validate the performance of the proposed prototype for RA-enabled wireless communication system. The experimental setup is shown in Fig. 2, where the 16-QAM is adopted and the distance between TX and RX is set to 4 meters (m). The system operates at a carrier frequency of 5.8 GHz with a data rate of 0.5 Mbps. Furthermore, the transmit power is set to 10 dBm. Given the bandwidth of 100 KHz, the tested system has a noise level of approximately -95 dBm. The servo adjusts the deflection angle of the directional antenna within a rotation range of \( [-\frac{\pi}{2},\frac{\pi}{2}] \). Moreover, the radar imaging map is employed to display the location of the RX.
	
\begin{figure}[!h]
	\centering
	\includegraphics[width=2.7in]{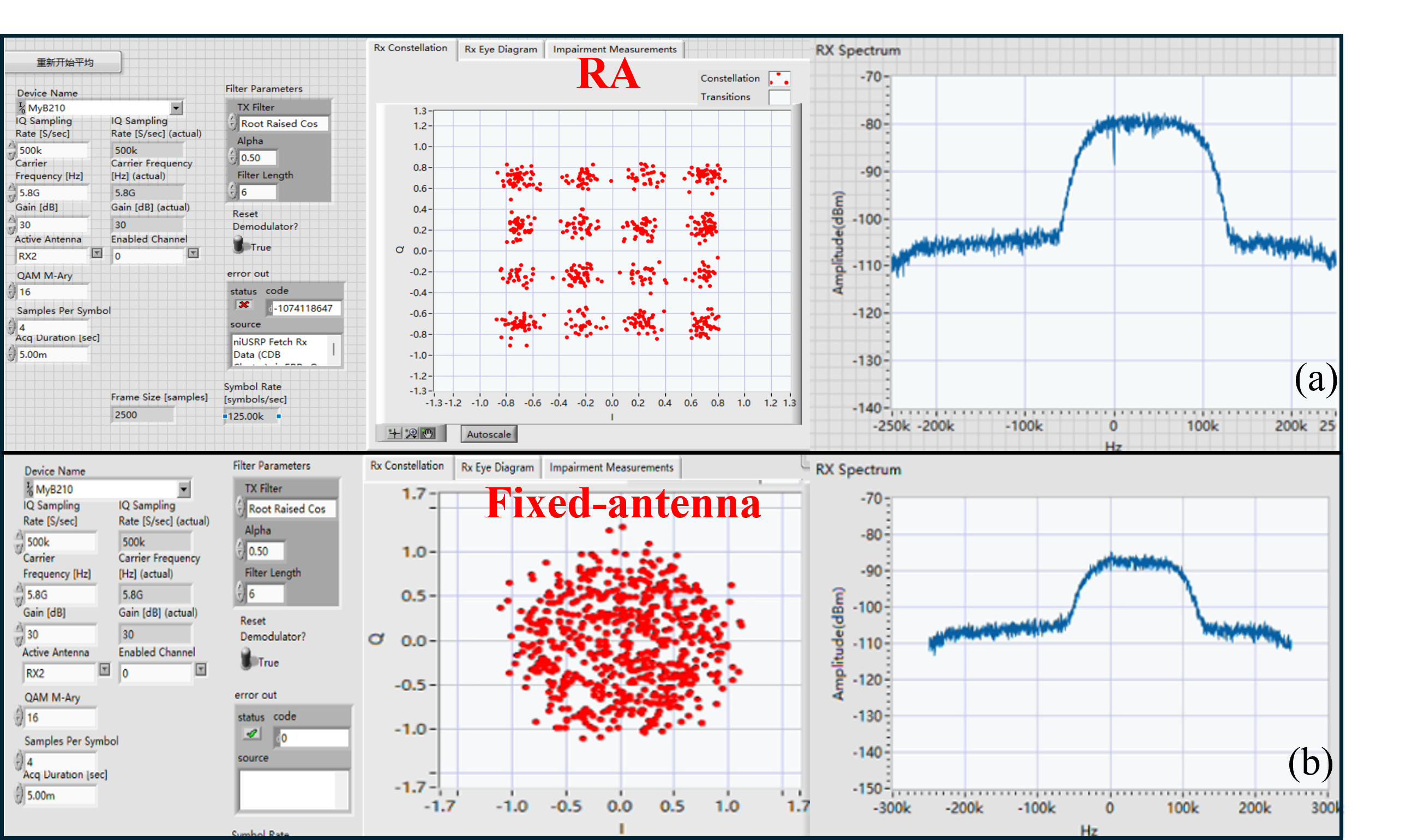} % 替换 "your-image-file" 为图片文件名，例如 "system_overview.png"
	\vspace{-0.15cm}
	\caption{Constellation graphs of (a) RA and (b) Fixed-antenna. }
	\label{fig:system}
	%\vspace{-0.15cm}
\end{figure}
	
	Fig. 3 shows the 16-QAM constellation diagrams of RA and fixed-antenna systems, where the RX's azimuth angle is set to  \( \frac{\pi}{3} \). Two interesting observations are made as follows. First, compared to the fixed-antenna system, the constellation diagram of the RA system is clearer and more regularly distributed, indicating a significant improvement in received signal quality. Second, we can observe that the RA system results in higher received signal power than the fixed-antenna counterpart, achieving an approximate gain of 7 dBm. This result indicates that the RA system exhibits superior communication performance to the fixed-antenna. 
	
		\begin{figure}[!h]
		\centering
		\vspace{-0.45cm}
		\includegraphics[width=3.1in]{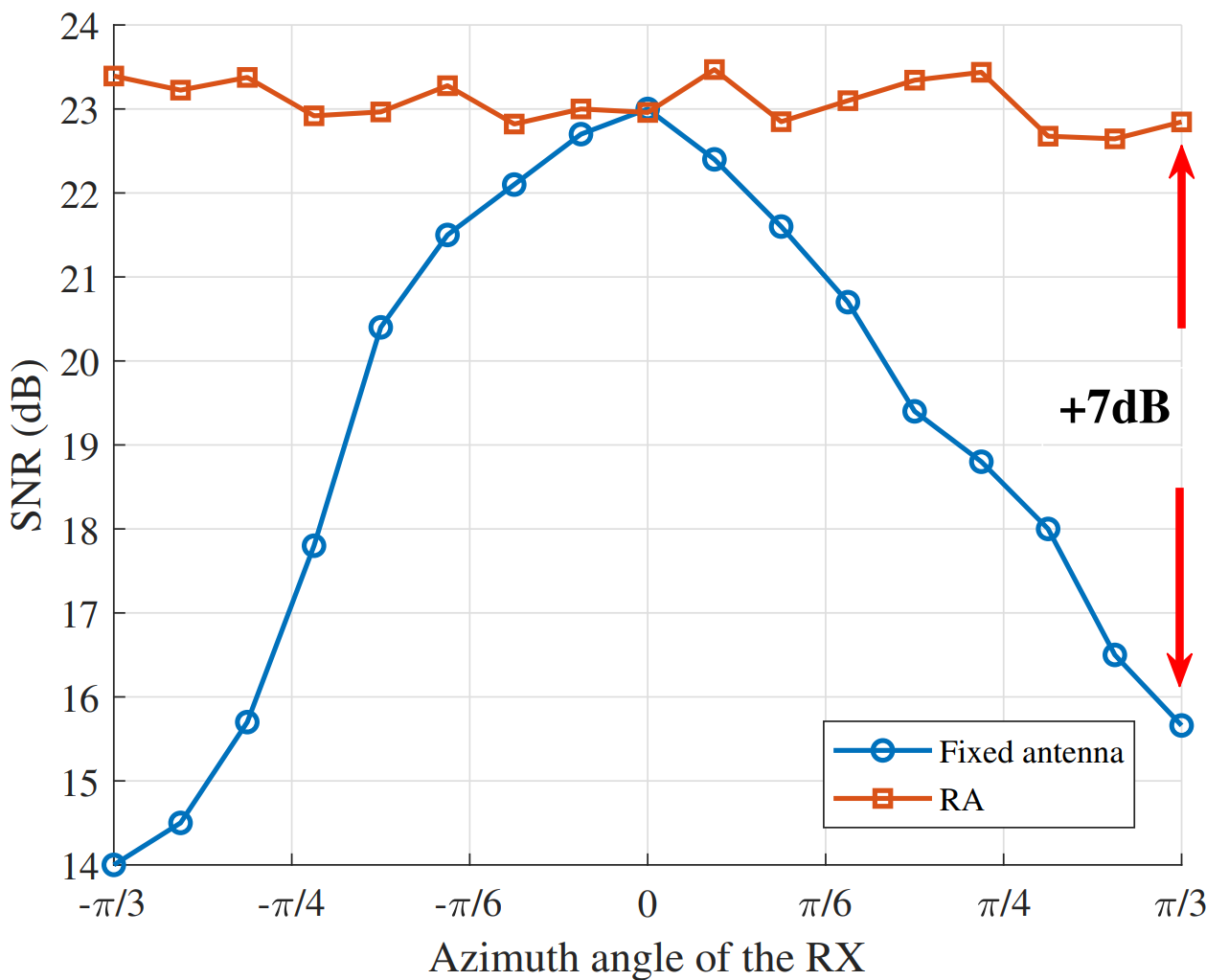} % 替换 "your-image-file" 为图片文件名，例如 "system_overview.png"
		%\vspace{-0.35cm}
		\caption{Received SNR versus the azimuth angle of the RX.}
		\label{fig:system}
		\vspace{-0.35cm}
	\end{figure}
	
	Fig. 4 illustrates the received signal-to-noise ratio (SNR) versus the RX's azimuth angle for both RA and fixed-antenna systems. As the RX's azimuth angle increases from 0 to \( \frac{\pi}{3} \) or decreases from 0 to \(-\frac{\pi}{3} \), the fixed-antenna system experiences a significant SNR degradation, which is expected since the fixed directional gain pattern can not be flexibly redirected toward the moving RX. In contrast, the RA system sustains a stable SNR level since it can dynamically reconfigure its gain pattern to constantly enhance the directional gain towards the RX direction. This result indicates that the RA can dynamically align its antenna boresight direction with the RX, ensuring robust communication performance even when the RX's location varies.
	
	\section{Conclusion}

	In this demonstration, we developed a prototype of radar sensing-aided RA communication system that integrates precise radar detection with dynamic antenna orientation. Experimental results indicated that our proposed RA communication system significantly outperforms that of the fixed-antenna counterpart in terms of received SNR. It was demonstrated that the RA has great potential in enhancing communication quality and expanding signal coverage, thereby validating its effectiveness and feasibility in dynamic wireless environments.

	\bibliographystyle{IEEEtran} % 使用IEEEtran样式
	
	\bibliography{references} % 指定.bib文件名称（无需扩展名）
	
	% 文档结束
\end{document}